\documentclass[a4paper,11pt]{article}

\usepackage[cp1250]{inputenc}

\newtheorem{dfn}{Definition}[section]
\newtheorem{tw}[dfn]{Theorem}

\newtheorem{rem}[dfn]{Remark}
\newtheorem{lem}[dfn]{Lemma}

\usepackage{anysize}
\usepackage{verbatim}
\usepackage{array}
\usepackage{enumerate}
\usepackage{amssymb} 
\usepackage{amsmath}
\usepackage{fancyhdr}
\usepackage{euscript}
\usepackage{latexsym}

\author{Micha\l \ Barski \\ \small  Faculty of Mathematics and Computer Science, University of Leipzig, Germany\\
\small Faculty of Mathematics, Cardinal Stefan Wyszy\'nski University in Warsaw, Poland
\\ \small{\it Michal.Barski@math.uni-leipzig.de}}

\title{\bf Quantile hedging on markets with proportional transaction costs}

\begin{document}


\maketitle
\renewcommand{\abstractname}{Abstract}
\begin{abstract}
In the paper a problem of risk measures on a discrete-time market
model with transaction costs is studied. Strategy effectiveness
and shortfall risk is introduced. This paper is a generalization
of quantile hedging presented in [4].
\end{abstract}
\begin{quote}
\end{quote}
\begin{quote}
\end{quote}
\begin{quote}
{\bf GEL Classification Numbers}: G11, G13
\end{quote}

\newcounter{Theorem}[section]
\newcounter{Proposition}[section]
\newcounter{Remark}[section]

\section{Introduction}
\quad \indent It is well known that on a classical market without
transaction costs the price $x_0$ of a contingent claim $C$ is
given as $x_0=\sup_{Q^\in\mathcal{Q}}\mathbf{E}^{Q}[C]$, where
$\mathcal{Q}$ is a set of all martingale measures equivalent to
the objective measure $P$. This means that if we have an initial
endowment $x\geq x_0$ then we can hedge $C$. Thus for $x$ there
exists a self-financing strategy $B$ for which the terminal value
$X_T^{x,B}$ is not smaller then $C$. If $x<x_0$ then we no longer
can hedge $C$. For each strategy $B$ we have $P(X_T^{x,B}<C)>0$.
The investor who wants to hedge $C$ in some way must consider some
risk connected with the fact that he is not able to hedge $C$
entirely. There appeared many risk measures introduced for
instance by Cvitani† and Karatzas [1], Pham [8], F\"{o}llmer and
Leukert [4] and [5]. Cvitanić and Karatzas study the following
risk measure:
$\inf_{B\in\mathcal{B}}\mathbf{E}[(C-X_T^{x,B})^{+}]$, where
$\mathcal{B}$ is a set of all self-financing strategies. Pham
introduced $L^p$ hedging in [8] and his risk measure is defined as
$\inf_{B\in\mathcal{B}}\mathbf{E}[l_p((C-X_T^{x,B})^{+})]$, where
$l_p(x)=\frac{x^p}{p}$. Another examples of risk measures are
provided
 by F\"{o}llmer and Leukert in [4]. They consider so called quantile hedging problem introducing
a random variable connected with the strategy $(x,B)$ by defining
\begin{gather*}
\varphi_{x,B}=\mathbf{1}_{\{X_T^{x,B}\geq C\}}+\frac{X_T^{x,B}}{C}
\mathbf{1}_{\{X_T^{x,B}< C\}}.
\end{gather*}
This random variable is called "success function" and its
expectation is an effectiveness measure connected with the
strategy $(x,B)$. Success function takes its values in the
interval $[0,1]$. If $(x,B)$ is a hedging strategy, then
$\varphi_{x,B}=1$, otherwise $P(\varphi_{x,B}<1)>0$ what implies
$\mathbf{E}[\varphi_{x,B}]<1$. Their aim is to find the strategy
$B$ to maximize $\mathbf{E}[\varphi_{x,B}]$ for a given $x$. In
the next paper [5] they also examine another risk measure which is
given as $\inf_{B\in\mathcal{B}}\mathbf{E}[l((C-X_T^{x,B})^+)]$,
where $l$ is a loss function.\\
\indent In this paper we study a problem of risk measures on
markets with proportional transaction costs. The main idea is
based on papers of F\"{o}llmer and Leukert on quantile hedging [4]
and minimizing shortfall risk [5]. On markets with transaction
costs we are given a multi-dimensional contingent claim $H$,
multi-dimensional wealth process $V_t^{v,B}$ and some cone $K_T$
which is constructed on a basis of transaction costs. The cone
$K_T$ indicates a partial ordering "$\succeq_{_T}$" in
$\mathbb{R}^{d}$ in the sense that $x\succeq_{_T}
y\Longleftrightarrow x-y\in K_T$. We say that strategy $(v,B)$
hedges $H$ if $V_T^{v,B}\succeq_{_T} H$. In papers [2], [6] and[7]
the authors provide characterization of the set
$\Gamma(H)\mathbb\subseteq{R}^d$ of initial endowments for which
there exists a hedging strategy $B$ such that
$V_T^{v,B}\succeq_{_T} H$. The problem arises, what in a sense, is
an optimal strategy for an initial endowment $v\notin\Gamma(H)$.
For the terminal wealth $V_T^{v,B}$ we introduce a set of
proportional transfers which is denoted by ${\EuScript
L}(V_T^{v,B},H)$. Simply speaking, for $L\in{\EuScript
L}(V_T^{v,B},H)$ we have $\frac{({V_T^{v,B}}_{\mid L})^i}{H^i}
=\frac{({V_T^{v,B}}_{\mid L})^j}{H^j}$ for all $i,j$ where
${V_T^{v,B}}_{\mid L}$ is a terminal wealth after a proportional
transfer $L$. For this ratio we denote $\frac{{V_T^{v,B}}_{\mid
L}}{H}:=\frac{({V_T^{v,B}}_{\mid L})^i}{H^i}$. In section $3$ we
introduce the "success function" which expectation is an
effectiveness measure of the strategy $(v,B)$ by setting
\begin{gather*}
\varphi_{v,B}=\mathbf{1}_{\{V_T^{v,B}\succeq_{_T}
H\}}+\underset{L\in{\EuScript
L}(V_T^{v,B},H)}{\text{ess}\sup}\frac{{V_T^{v,B}}_{\mid
L}}{H}\mathbf{1}_{\{V_T^{v,B}\succeq_{_T} H\}^c}.
\end{gather*}
We establish some useful properties of the success function. It
appears that $\varphi_{v,B}\in[0,1]$ and if $v\in\Gamma(H)$ then
$\varphi_{v,B}=1$ for the hedging strategy $B$, whereas for
$v\notin\Gamma(H)$ we have $P(V_T^{v,B}<1)>0$ for each strategy
$B$ what implies $\mathbf{E}[\varphi_{v,B}]<1$. Our aim is to find
the strategy $B$ for the initial endowment $v$ to maximize
$\mathbf{E}[\varphi_{v,B}]$. We consider also another problem. For
$1\geq\varepsilon\geq 0$ we characterize the set
$\Gamma_{\varepsilon}(H)\subseteq \mathbb{R}^d$ of initial
endowments for which there exists the strategy $B$ such that
$\mathbf{E}[\varphi_{v,B}]\geq 1-\varepsilon$. These are two
aspects of quantile hedging which are analogous to problems
presented by F\"{o}llmer and
Leukert.\\
Then, in section $6$, we introduce a shortfall risk in quantile
hedging. Shortfall is defined as
\[
s(V_T^{v,B})=
\begin{cases}
\qquad\qquad 0 \quad &\text{on the set}\quad \{V_T^{v,B}\succeq_{_T}H\}\\
\Big( 1-\underset{L\in{\EuScript
L}(V_T^{v,B},H)}{\text{ess}\sup}\frac{{V_T^{v,B}}_{\mid
L}}{H}\Big) \quad &\text{on the set}\quad
{\{V_T^{v,B}\succeq_{_T}H\}}^c
\end{cases}.
\]
Shortfall is a $[0,1]$-valued random variable which is equal to
$0$ if $V_T^{v,B}\succeq_{_T}H$ and it is strictly positive if
$V_T^{v,B}\nsucceq_{_T}H$. It describes the part of the contingent
claim which is not hedged by the strategy $(v,B)$. We study the
problem of minimizing shortfall risk given as
$\mathbf{E}[u(s(V_T^{v,B}))]$, where
$u:[0,1]\longrightarrow\mathbb{R}$ is a loss function. We accept
here the assumption, that the investor considers only the
percentage of the contingent claim which is not hedged as a loss,
not the value of this part. As before, we study two problems.
Firstly, in section $6$, we characterize the strategy $B$ which
minimizes shortfall risk. Secondly, in section~$7$, we
characterize the set $\Gamma^u_\alpha(H)$ of initial endowments
for which there exists the strategy $B$ such that
$\mathbf{E}[u(s(V_T^{v,B}))]\leq
\alpha$ for a given number $\alpha\geq 0$.\\
\indent In section $8$ we show how F\"{o}llmer's and Leukert's
theory can be obtained under zero transaction costs. Since
condition ${\bf EF}$ imposed in [6] is not satisfied, we use
results shown in [2].

\section{Market with proportional transaction costs}
\indent In this section we present some results obtained by
Kabanov, R\'{a}sonyi, Stricker in papers [6] and [7] which deal
with conditions for the absence of arbitrage under friction. We
particularly need a hedging theorem providing description of the
set of
initial endowments which allow to hedge the contingent claim.\\
\indent Let
$(\Omega,\mathcal{F},(\mathcal{F}_{t})_{t=0,1,...T},P)$ be a
probability space equipped with a complete, discrete-time
filtration. We assume that $\mathcal{F}_{0}$ is a trivial
$\sigma$-field and that $\mathcal{F}_T=\mathcal{F}$. On $\Omega$
we are given a strictly positive $\mathbb{R}^{d}$-valued, adapted
process $S_t$ which describes the prizes of $d$ traded securities.
We can assume that, for instance the first component is a price of
a bond, but it is not necessary for further consideration.
Proportional transaction costs are given as the process
$\Lambda_t=(\lambda_t^{i,j})_{i,j=1,2,...,d}$ with values in the
set $M^d_+$ of matrices with non-negative, adapted entries and
zero diagonal. If we want to increase the $j$-th stock account by
the amount $L^{ij}\geq 0$ at time $t$, then we have to transfer an
amount $(1+\lambda^{ij}_t)L^{ij}$ from the $i$-th account. The
quantity $\lambda^{ij}_tL^{ij}$ is lost because of occurring
transaction costs. Given an initial endowment $v\in\mathbb{R}^d$
we invest in stocks at each time $t=0,1,...,T$. The agent's
position at time $t$ can be described either by vector
$\widehat{V}_t$ of stock units or by vector $V_t$ of values
invested in each stock. The relation between these quantities is:
$V^i_t=\hat{V}^i_tS^i_t$. Operator $"\widehat\quad"$  will be used
also for any random vector $Z$ and $\widehat{Z}$ stands for
$(\frac{Z^1}{S^1},...,\frac{Z^d}{S^d})$. A self-financing
portfolio is defined by its increments as follows
\begin{gather*}
\Delta V_t^i=\hat{V}^i_{t-1}\cdot\Delta S^i_t+\Delta B^i_t \quad
i=1,...,d,\quad t=0,1,...,T,
\end{gather*}
with convention for initial values $V^i_{-1}=v^i, \ S_{-1}=S_0, \
L^{ij}_{-1}=0$ for $i,j=1,2,...,d$ and where
\begin{gather*}
\Delta B^i_t:=\sum_{j=1}^{d}\Delta
L^{ji}_{t}-\sum_{j=1}^{d}(1+\lambda^{ij}_{t})\Delta L_{t}^{ij}.
\end{gather*}
Here we denote $\Delta Y_t=Y_t-Y_{t-1}$ for each process $Y$. The
adapted, increasing and non-negative process $L^{ij}$ represents
the net cumulative transfers from the position $i$ to the position
$j$ under transaction costs. The increment $\Delta V_t^i$ of value
on $i$-th stock account consists of two parts: the increment
$\hat{V}_t^i\Delta S_t^i$ due to the price movements and the
increment $\Delta B_t^i$ caused by agent's action at time $t$.
Since the pair $(v,B)$ determines the wealth process $V_t^{v,B}$,
we will treat it as a trading strategy.\\
\indent In the sequel we will use the following notation :
$L^0(A,\mathcal{F}_t)$, where $A\subseteq\mathbb{R}^d$ is a set of
$\mathcal{F}_t$ measurable random variables which take values in
the set $A$. $L^0(M^d_+,\mathcal{F}_t)$ stands for matrices which
entries are non-negative and $\mathcal{F}_t$ measurable random
variables. Let
\begin{align*}
M_t(\omega):=\Big\{ x\in\mathbb{R}^d : \quad &\exists \ L\in
L^0(M^d_+,\mathcal{F}_t) \quad \text{such
that}\\[2ex]
&
x^i=\sum_{j=1}^{d}(1+\lambda_{t}^{ij}(\omega))L^{ij}(\omega)-\sum_{j=1}^{d}L^{ji}(\omega)
\Big\}
\end{align*}
be a set of position which can be converted into zero by a
non-negative transfer. This set is a polyhedral cone. Let
$K_t:=\mathbb{R}^d_++M_t$ and $F_t:=K_t\cap(-K_t)$. The set $K_t$,
which is called the solvency region, is a polyhedral cone. It is
formed by vectors which can be transformed into a vector with only
non-negative components by a positive transfer, thus by adding a
vector from $-M_t$. $F_t$ represents positions which can be
converted into zero and vice versa. $F_t$ is a linear
space.\\
\indent We shall say that a strategy $(0,B)$ is a weak arbitrage
opportunity at time $t$ if $V_{t}^{0,B}\in K_t$ and
$P(V_t^{0,B}\in K_t \setminus F_t)>0$. There is an absence of a
weak arbitrage opportunity if there does not exist arbitrage
opportunity at any time. The absence of a weak arbitrage
opportunity ({\it strict no-arbitrage property}) can be expressed
in geometric terms:
\begin{gather*}
{\bf NA}^{s} : \qquad R_t\cap L^{0}(K_t,\mathcal{F}_t)\subseteq
L^{0}(F_t,\mathcal{F}_t) \quad \text{for} \ t=0,1,...,T,
\end{gather*}
where
\begin{gather*}
R_t:=\Big\{V_{t}^{0,B} : B\in\mathcal{B}; \ \mathcal{B}\text{-set
of all strategies}\Big\}.
\end{gather*}
The set $R_t$ describes wealth at time $t$ which can be obtained
starting with the zero initial endowment.\\
\noindent Let us define an {\it efficient friction} condition.
\begin{gather*}
{\bf EF}:\qquad \text{The cones} \ K_t(\omega) \ \text{are proper,
i.e.} \ F_t(\omega)=\{0\} \ \text{for each} \ (\omega,t).
\end{gather*}
Under {\bf EF} the condition ${\bf NA}^{s}$ can be rewritten as
$R_t\cap L^0(K_t,\mathcal{F}_t)=\{0\} \ \text{for}\ t=0,1,...,T$.
Under {\bf EF} there are some equivalent conditions to ${\bf
NA}^{s}$.
For more details see [6].\\
\indent The most important result for this paper is a description
of the set of initial endowments which allow to hedge the
contingent claim. Let us start with the fact that the cone $K_t$
generates a partial ordering "$\succeq_{_t}$" on $\mathbb{R}^d$ in
the sense that $x\succeq_{_t}y\Longleftrightarrow x-y\in K_t$.
Contingent claim $H$ is an $\mathbb{R}^d$ valued random variable
and the set
\begin{gather*}
\Gamma(H):=\{v\in\mathbb{R}^d : \text{there exists a strategy} \ B
\ \text{such that} \ V_T^{v,B}\succeq_{_T}H\}.
\end{gather*}
stands for all hedging initial endowments. For simplicity we
assume that $H\succeq_{_T} c\mathbf{1}$ for some $c\in\mathbb{R}$.
The next theorem presented in [7] provides description of the set
$\Gamma(H)$. \vskip 10pt
\begin{tw}
Assume that {\bf EF} and ${\bf NA}^s$ are satisfied. Then
\begin{gather*}
\Gamma(H)=\Big\{v\in\mathbb{R}^d :
\hat{Z}_0v\geq\mathbf{E}\hat{Z}_TH \quad \forall
Z\in\mathcal{Z}\Big\}
\end{gather*}
where $\mathcal{Z}$ is the set of bounded martingales such that
$\hat{Z}_t\in L^0(K_t^\ast,\mathcal{F}_t)$ for $t=0,1,...,T$ and
where $K_t^\ast$ denotes the dual cone to the cone $K_t$.
\end{tw}
From now on we assume that conditions {\bf EF} and ${\bf NA}^s$
are satisfied.

\section{Strategy effectiveness }
In this section we introduce a success function $\varphi_{v,B}$
for the strategy $(v,B)$ and establish its properties. Its
expectation under $P$ is in fact some kind of risk measure, but
more adequate risk measure will be defined in section $6$. This
one we accept rather as an effectiveness
measure.\\
We will consider only admissible strategies and from now on we
assume that $H\succeq_{_T}0$ almost everywhere.
\begin{dfn}
Strategy $(v,B)$ is admissible if $V_T^{v,B}\succeq_{_T}0$.
\end{dfn}
Let $(v,B)$ be an admissible strategy. Our aim is to describe its
effectiveness regarding the contingent claim $H$. Divide $\Omega$
into two parts: $\{V_T^{v,B}\succeq_{_T}H\}$ and
$\{V_T^{v,B}\succeq_{_T}H\}^c$. On the set
$\{V_T^{v,B}\succeq_{_T}H\}$ we put $\varphi_{v,B}=1$. The next
part of this section is to define $\varphi_{v,B}$ on the set
$\{V_T^{v,B}\succeq_{_T}H\}^c$
and examine its basic properties.\\
For the terminal wealth $V_T^{v,B}$ and transfer $L\in
L^0(M^d_+,\mathcal{F}_T)$ we will consider $V_T^{v,B}$ after
transfer $L$ under transaction costs at time $T$ given by
\begin{gather*}
({V_T^{v,B}}_{\mid
L})^i=(V_T^{v,B})^i+\sum_{j=1}^{d}L^{ji}-\sum_{j=1}^{d}(1+\lambda^{ij}_T)L^{ij}.
\end{gather*}
\noindent
 In the set of all transfers $L^0(M^d_+,\mathcal{F}_T)$
we distinguish a subclass of proportional transfers.

\begin{dfn}
Assume that for an admissible strategy $(v,B)$ holds
$V_T^{v,B}\nsucceq_{_T} H$. Transfer $L \in
L^0(M^d_+,\mathcal{F}_T)$ is a proportional transfer if there
exists $c_L\in L^0(\mathbb{R},\mathcal{F}_T)$ such that
\begin{gather*}
{V_T^{v,B}}_{\mid L}=c_L \cdot H.
\end{gather*}
\end{dfn}
${\EuScript L}(V_T^{v,B},H)$ stands for the class of all
proportional transfers and for $L\in{\EuScript L}(V_T^{v,B},H)$ we
denote $\frac{{V_T^{v,B}}_{\mid L}}{H}:=c_L$.
\begin{rem}
${\EuScript L}(V_T^{v,B},H)$ is not empty since $(v,B)$ is
admissible. This means that there exists $L_0\in
L^{0}(M^d_+,\mathcal{F}_T)$ for which ${V_T^{v,B}}_{\mid
L_{0}}=0$, thus ${c_L}_0=\frac{{V_T^{v,B}}_{\mid L_0}}{H} =0.$
\end{rem}
The meaning of the class of proportional transfers is to achieve
the same "rate of hedge" on each stock account. We want to make
this rate as high as possible. Thus on the set
$\{V_T^{v,B}\succeq_{_T}H\}^c$ we define $\varphi_{v,B}$ as
$\underset{L\in{\EuScript L}(V_T^{v,B},H)}{\text{ess}
\sup}\frac{{V_T^{v,B}}_{\mid L}}{H}$. This leads to the following
definition of the success function~:
\begin{gather*}
\varphi_{v,B}=\mathbf{1}_{\{V_T^{v,B}\succeq_{_T}H\}}+
\underset{L\in{\EuScript L}(V_T^{v,B},H)}{\text{ess}
\sup}\frac{{V_T^{v,B}}_{\mid L}}{H}
 \ \mathbf{1}_{\{V_T^{v,B}\succeq_{_T}H\}^c}.
\end{gather*}
\begin{lem}
Assume that $V_T^{v,B}\nsucceq_{_T}H$. There exists an optimal
transfer $\widehat{L}\in{\EuScript L}(V_T^{v,B},H)$ such that
\begin{gather*}
\underset{L\in{\EuScript L}(V_T^{v,B},H)} {\emph{ess}\sup}
\frac{{V_T^{v,B}}_{\mid L}}{H}=
\frac{{V_T^{v,B}}_{\mid\widehat{L}}}{H}.
\end{gather*}
\end{lem}
{\it Proof}\\
Let us consider two geometrical objects which depend on $\omega$:
the translated polyhedral cone $V_T^{v,B}+(-M_T)$ with its
boundary $\partial(V_T^{v,B}+(-M_T))$ and the line spanned by the
vector $H$. $V_T^{v,B}+(-M_T)$ is generated by $m$ measurable
vectors $\xi_1,\xi_2,...,\xi_m$, where $d\leq m\leq d(d-1)$ and
can be represented as an intersection of $l$ half-spaces for some
$l$. The $i$-th half-space is spanned by $d-1$ generators
$\xi_{i_1},\xi_{i_2},...,\xi_{i_{d-1}}$ from the set
$\xi_1\xi_2,...,\xi_m$. Putting
$g_i=\xi_{i_1}\times\xi_{i_2}\times...\times\xi_{i_{d-1}}$ where
$\times$ denotes the cross product, we obtain a measurable vector
which is orthogonal to each vector from the set
$\xi_{i_1},\xi_{i_2},...,\xi_{i_{d-1}}$. Thus the $i$-th
half-space has the following representation:
\begin{gather*}
\Big\{x\in\mathbb{R}^d :\quad (x-V_T^{v,B})\cdot g_i\geq 0\Big\},
\end{gather*}
and the boundary of the cone can be represented as:
\[
x\in\partial(V_T^{v,B}+(-M_T))\Longleftrightarrow
\begin{cases}
(x-V_T^{v,B})\cdot g_i\geq 0 \quad\forall i=1,2,...,l\\
(x-V_T^{v,B})\cdot g_i=0 \quad\text{for some} \  i=1,2,...,l.
\end{cases}
\]
On the other hand the line spanned by the vector $H$ can be
represented as
\begin{gather*}
x\in span\{H\}\Longleftrightarrow x\cdot h_i=0 \quad \forall
i=1,2,...,d-1,
\end{gather*}
where $\{H,h_1,h_2,...,h_{d-1}\}$ is a basis in $\mathbb{R}^d$,
each vector $h_i$ is measurable and $H\bot h_i$ for all
$i=1,2,...,d-1$. Such basis can be obtained by taking the set
$\{H,H+e_1,H+e_2...,H+e_d\}$, where $\{e_1,e_2,...,e_d\}$ is a
standard basis in $\mathbb{R}^d$, choosing a subset of $d$ linear
independent vectors containing $H$ and then orthogonalizing it
starting with the vector
$H$.\\
There exists exactly one positive point $\widehat{V}$ of
intersection $\partial(V_T^{v,B}+(-M_T))$ with $span\{H\}$. Since
it is a solution of linear system with measurable coefficients
\[
\begin{cases}
(x-V_T^{v,B})\cdot g_i\geq 0 &\forall i=1,2,...,l\\
(x-V_T^{v,B})\cdot g_i=0 &\text{for some} \  i=1,2,...,l\\
x\cdot h_i=0 &\forall i=1,2,...,d-1,
\end{cases}
\]
it is a measurable random vector. Hence also measurable is
$\widehat{c}$,
where $\widehat{V}=\widehat{c}H$.\\
Each transfer is represented by adding to $V_T^{v,B}$ some vector
from the cone $(-M_T)$. As $\widehat{L}$ we get the transfer
represented by $\widehat{V}-V_T^{v,B}$. From construction of
$\widehat{V}$ we conclude that for any other proportional transfer
such that ${V_T^{v,B}}_{\mid L}=\bar{c}H$ we have
$\bar{c}\leq\widehat{c}$. As a consequence we obtain
\begin{gather*}
\widehat{c}=\underset{L\in{\EuScript L}(V_T^{v,B},H)}
{\text{ess}\sup} \frac{{V_T^{v,B}}_{\mid L}}{H}=
            \frac{{V_T^{v,B}}_{\mid\widehat{L}}}{H}.
\end{gather*}
\quad\hfill$\square$

\begin{rem}
The success function fulfils
\begin{gather*}
0\leq\varphi_{v,B} \ \mathbf{1}_{{\{V_T^{v,B}\succeq_{_T}H\}}^c}
<1
\end{gather*}
\end{rem}
{\it Proof}\\
$\varphi_{v,B} \ \mathbf{1}_{{\{V_T^{v,B}\succeq_{_T}H\}}^c}\geq
0$ since $(v,B)$ is admissible. If $\varphi_{v,B} \
\mathbf{1}_{{\{V_T^{v,B}\succeq_{_T}H\}}^c} \geq1$ then
$\mathbf{1}_{{\{V_T^{v,B}\succeq_{_T}H\}}^c}
\frac{{V_T^{v,B}}_{\mid \widehat{L}}}{H}\geq 1$. This implies
${V_T^{v,B}}_{\mid \hat{L}}\geq H$ on the set
$\{V_T^{v,B}\succeq_{_T}H\}^c$ but this means that
${V_T^{v,B}}\succeq_{_T}H$ what is a contradiction.
\hfill$\square$\\ \\
To summarize, the success function $\varphi_{v,B}$ is equal to $1$
if
 $V_T^{v,B}\succeq_{_T} H$ and strictly smaller then $1$ if
$V_T^{v,B}\nsucceq_{_T}H$.\\ \\
In the next part of the paper we will work with the set
\begin{gather*}
\mathcal{R}:=\left\{\varphi: \ 0 \leq\varphi\leq 1; \ \varphi \
\text{is} \ \mathcal{F}_T \ \text{measurable}\right\}
\end{gather*}
of $\mathcal{F}_T$ measurable functions which takes values in
$[0,1]$. \vskip 10pt \noindent We start with two useful properties
of the success function. \vskip 10pt \noindent
\begin{lem}
Assume that $(v,B)$ is an admissible strategy. Then
$v\in\Gamma(H\varphi_{v,B})$.
\end{lem}
{\it Proof}\\
In view of lemma 3.4 we have
\begin{align*}
H\varphi_{v,B}&=H \ \mathbf{1}_{\{V_T^{v,B}\succeq_{_T}H\}}+ H
\underset{L\in{\EuScript L}(V_T^{v,B},H)}{\text{ess} \
\sup}\frac{{V_T^{v,B}}_{\mid L}}{H}
\ \mathbf{1}_{\{V_T^{v,B}\succeq_{_T}H\}^c}\\[2ex]
&=H \ \mathbf{1}_{\{V_T^{v,B}\succeq_{_T}H\}}+{V_T^{v,B}}_{\mid
\widehat{L}}
 \ \mathbf{1}_{\{V_T^{v,B}\succeq_{_T}H\}^c}\\[2ex]
&{_{_T}}\!\!\!\preceq  V_T^{v,B}
\end{align*}
where $\widehat{L}$ is an optimal proportional transfer. Thus we
have $v\in\Gamma(H\varphi_{v,B})$.
\hfill$\square$\\ \\
\begin{lem}
Assume that $(v,B)$ is a hedging strategy for a modified
contingent claim $H\varphi$ for some function
$\varphi\in\mathcal{R}$. Then $\varphi_{v,B}\geq \varphi$.
\end{lem}
{\it Proof}\\
Since $V_T^{v,B}\succeq_{_T} H\varphi$, there exists transfer
$M\in L^0(M^d_+,\mathcal{F}_T)$ such that ${V_T^{v,B}}_{\mid
M}-H\varphi\geq 0$. Let $N\in{\EuScript L}({V_T^{v,B}}_{\mid
M}-H\varphi,H)$ be any proportional transfer on the set
$\{V_T^{v,B}\succeq_{_T}H\}^c$ such that
$\frac{{\big({V_T^{v,B}}_{\mid M}-H\varphi\big)}_{\mid
N}}{H}=\gamma$ for some $\gamma\geq 0$. Let us consider the
terminal wealth $V_T^{v,B}$ on the set
$\{V_T^{v,B}\succeq_{_T}H\}^c$ after transfer $K$ described as
follows: first change $V_T^{v,B}$ by the transfer $M$ and then
change ${V_T^{v,B}}_{\mid M}-H\varphi$ by transfer $N$. The
terminal wealth $V_T^{v,B}$ after transfer $K$ is thus given as
\begin{gather*}
{V_T^{v,B}}_{\mid K}=H\varphi+({V_T^{v,B}}_{\mid
M}-H\varphi)_{\mid N}.
\end{gather*}
It is clear that $K\in{\EuScript L}(V_T^{v,B},H)$ since
\begin{gather*}
{V_T^{v,B}}_{\mid K}=H\varphi+H\gamma=(\varphi+\gamma)H.
\end{gather*}
This leads to the following inequalities
\begin{align*}
\varphi_{v,B}&=\mathbf{1}_{\{V_T^{v,B}\succeq_{_T}H\}}+
                 \underset{L\in{\EuScript L}(V_T^{v,B},H)}{\text{ess} \sup}\frac{{V_T^{v,B}}_{\mid L}}{H}
                 \mathbf{1}_{\{V_T^{v,B}\succeq_{_T}H\}^c}\\[2ex]
&\geq \mathbf{1}_{\{V_T^{v,B}\succeq_{_T}H\}}+
\frac{{V_T^{v,B}}_{\mid K}}{H}
\mathbf{1}_{\{V_T^{v,B}\succeq_{_T}H\}^c}\\[2ex]
&=\mathbf{1}_{\{V_T^{v,B}\succeq_{_T}H\}}+
(\varphi+\gamma) \ \mathbf{1}_{\{V_T^{v,B}\succeq_{_T}H\}^c}\\[2ex]
&\geq \varphi.
\end{align*}
\hfill$\square$\\

\section{Quantile hedging - effectiveness maximization}
The set $\Gamma(H)$ is a set of all initial endowments which allow
to hedge the contingent claim $H$. If $v\in\Gamma(H)$ then there
exists a strategy $B\in\mathcal{B}$ such that
$V_T^{v,B}\succeq_{_T}H$. Suppose that we are given an initial
capital $v_0$, such that $v_0\notin\Gamma(H)$. A natural question
arises : what is an optimal strategy for $v_0$ ? As the optimality
criteria we accept an expectation of the success function under
measure $P$. If for two admissible strategies $(v,B)$ and
$(\bar{v},\bar{B})$ holds $\mathbf{E}[\varphi_{v,B}]\geq
\mathbf{E}[\varphi_{\bar{v},\bar{B}}]$ then strategy $(v,B)$ is at
least as effective as $(\bar{v},\bar{B})$. If $(v,B)$ is at least
as effective as any other admissible strategy, then it is called
optimal. The problem of finding optimal strategy for $v_0$ is a
first aspect of quantile hedging problem and we formally formulate
it as follows :
\begin{quote}
{\it For a fixed initial endowment $v_0\in\Gamma(0)$ such that
$v_0\notin \Gamma(H)$ find an admissible strategy $(v,B)$, where
$v_0\succeq_{_0} v$, such that
$\mathbf{E}[\varphi_{v,B}]\longrightarrow \max$. }
\end{quote}
To describe optimal strategy, we start with the following theorem.
\vskip 10pt
\begin{tw}
There exists a function $\tilde{\varphi}\in\mathcal{R}$ which is a
solution of the problem
\begin{align*}
\mathbf{E}[\varphi]&\longrightarrow\max\\[2ex]
v_0&\in\Gamma(H\varphi).
\end{align*}
\end{tw}
{\it Proof}\\
Let us denote $\mathcal{R}_0:=\{\varphi\in\mathcal{R} :
v_0\in\Gamma(H\varphi)\}$. $\mathcal{R}_0\neq\emptyset$ since
$0\in\mathcal{R}_0$. Let $\varphi_n\in\mathcal{R}_0$ be a sequence
of elements such that
$\mathbf{E}[\varphi_n]~\longrightarrow~\sup_{\varphi\in\mathcal{R}_0}\mathbf{E}[\varphi]$.
Since $\{\varphi_n\}$ is a sequence of elements from a hull in
$L^{\infty}(\Omega)$, there exists a subsequence $\varphi_{n_k}$
which converges to $\tilde{\varphi}$ in a weak $\ast$ topology.
One can prove that $\tilde{\varphi}$ belongs to $\mathcal{R}$. We
will show that $v_0\in\Gamma(H\tilde{\varphi})$.
 Each element of the sequence $\{\varphi_n\}$ satisfies
$\widehat{Z}_0 v_0\geq\mathbf{E}[\widehat{Z}_T H\varphi_n] \quad
\forall Z\in\mathcal{Z}$, and $\tilde{\varphi}$ as a weak limit
satisfies
\begin{gather*}
\forall Z\in\mathcal{Z}\qquad \widehat{Z}_0
v_0\geq\mathbf{E}[\widehat{Z}_T
H\varphi_{n_k}]\underset{k}{\longrightarrow}\mathbf{E}[\widehat{Z}_TH\tilde{\varphi}].
\end{gather*}
Thus $v_0\in\Gamma(H\tilde{\varphi})$. \hfill$\square$
\\ \\
The next theorem provides the solution of our problem. \vskip 10pt
\begin{tw}
Let $\tilde{\varphi}$ be a function from theorem $4.1$, and the
strategy $(v_0,B)$ be a hedging strategy for the modified
contingent claim $H\tilde{\varphi}$. Then $(v_0,B)$ is an optimal
strategy. Furthermore, $\tilde{\varphi}=\varphi_{v_0,B}$.
\end{tw}
{\it Proof}\\
$(v_0,B)$ is admissible since $V_T^{v_0,B}\succeq_{_T}H\tilde{\varphi}\succeq_{_T}0$.\\
Let $(\bar{v},\bar{B})$ be any admissible strategy such that
$v_0\succeq_{_0}\bar{v}$. Then by lemma 3.6 we have:
$\bar{v}\in\Gamma(H\varphi_{\bar{v},\bar{B}})$ and this implies
that $v_0\in\Gamma(H\varphi_{\bar{v},\bar{B}})$. From theorem
$4.1$ we have
\begin{equation}
\mathbf{E}[\varphi_{\bar{v},\bar{B}}]\leq
\mathbf{E}[\tilde{\varphi}].
\end{equation}
Now, let us consider the strategy $(v_0,B)$. Since
$V_T^{v_0,B}\succeq_{_T} H\tilde{\varphi}$, by lemma 3.7 we have:
\begin{equation}
\varphi_{v_0,B}\geq \tilde{\varphi}.
\end{equation}
By virtue of $(4.2.1)$ and $(4.2.2)$ we have
$\varphi_{v_0,B}=\tilde{\varphi}$. Hence $(v_0,B)$ is optimal.
\hfill$\square$

\section{Quantile hedging - sets with a fixed level of effectiveness}
Assume, that we are given a number $\varepsilon\in[0,1]$. We want
to characterize strategies which effectiveness is not smaller then
$1-\varepsilon$. This is the second aspect of quantile hedging and
in fact our task is to characterize the set
$\Gamma_\varepsilon(H)$ which is given as
\begin{align*}
\Gamma_{\varepsilon}(H)=\Big\{v\in\mathbb{R}^{d}: \text{there
exists an \ }& \text{admissible strategy}
\ B \\[2ex]
&\text{such that \ } \mathbf{E}[\varphi_{v,B}]\geq 1-\varepsilon
\Big\}.
\end{align*}
It is clear that $\Gamma_{\varepsilon_1}(H)\subseteq
\Gamma_{\varepsilon_2}(H)$ if $\varepsilon_1\leq\varepsilon_2$.
Hence set $\Gamma_\varepsilon(H)$ contains the set
$\Gamma(H)=\Gamma_0(H)$, for any $\varepsilon\in[0,1]$ but it can
contain more elements as the initial capitals which allow to hedge
$H$ with some loss of effectiveness.\\
Let us set
\begin{gather*}
\mathcal{M}:=\{\varphi\in\mathcal{R} : \quad
\mathbf{E}[\varphi]\geq 1-\varepsilon\}.
\end{gather*}
The next theorem provides a description of the set
$\Gamma_\varepsilon(H)$. \vskip 10pt
\begin{tw}
The set $\Gamma_\varepsilon(H)$ admits the following
representation
\begin{gather*}
\Gamma_{\varepsilon}(H)=\bigcup_{\varphi\in\mathcal{M}}\Gamma(H\varphi).
\end{gather*}
\end{tw}
{\it Proof}\\
$\subseteq$\\
Let $v\in\Gamma_{\varepsilon}(H)$. Then there exists
$B\in\mathcal{B}$ such that $V_T^{v,B}\succeq_{_T}0$ and
$\mathbf{E}[\varphi_{v,B}]\geq 1-\varepsilon$. Thus
$\varphi_{v,B}\in\mathcal{M}$ and
\begin{gather*}
\Gamma(H\varphi_{v,B})\subseteq
\bigcup_{\varphi\in\mathcal{M}}\Gamma(H\varphi).
\end{gather*}
But $v\in\Gamma(H\varphi_{v,B})$ by lemma 3.6, and thus
$v\in\bigcup_{\varphi\in\mathcal{M}}\Gamma(H\varphi)$.
\\
$\supseteq$\\
Let $v\in\bigcup_{\varphi\in\mathcal{M}}\Gamma(H\varphi)$. Then
there exists $\varphi\in\mathcal{M}$ such that
$v\in\Gamma(H\varphi)$. Let us consider the strategy $(v,B)$ which
hedges the modified contingent claim $H\varphi$. Then by lemma 3.7
we have
\begin{gather*}
V_T^{v,B}\succeq_{_T}H\varphi\quad\Longrightarrow\quad\varphi_{v,B}\geq
\varphi,
\end{gather*}
and as a consequence
$\mathbf{E}[\varphi_{v,B}]\geq\mathbf{E}[\varphi]\geq
1-\varepsilon$. Finally, we have $v\in\Gamma_\varepsilon(H)$.
\hfill$\square$

\section{Risk measure in quantile hedging - minimizing shortfall risk}
On markets without transaction costs shortfall is defined as
$(C-X_T^{x,B})^+$, where $a^+=\max\{a,0\}$. In this section we
introduce a shortfall connected with the strategy $(v,B)$ under
transaction costs. To this end we use the set of proportional
transfers. Shortfall risk is introduced as an expectation of a
loss function of shortfall. Our aim is to minimize shortfall risk
for a fixed initial capital over all admissible strategies.
\\
In section $3$ we introduced a random variable
$\underset{L\in{\EuScript
L}(V_T^{v,B},H)}{\text{ess}\sup}\frac{{V_T^{v,B}}_{\mid L}}{H}$
defined on the set $\{V_T^{v,B}\succeq_{_T}H\}^c$. It describes
the part of the contingent claim which is successfully hedged. As
shortfall we accept the remaining part:
$\Big(1-\underset{L\in{\EuScript
L}(V_T^{v,B},H)}{\text{ess}\sup}\frac{{V_T^{v,B}}_{\mid
L}}{H}\Big)$. Let us start with formal definition.
\begin{dfn}
A shortfall of an admissible strategy $(v,B)$ is a random variable
set as
\[
s\left(V_T^{v,B}\right)=
\begin{cases}
\qquad\qquad0 &\emph{on the set} \ \left\{V_T^{v,B}\succeq_{_T}H\right\}\\[2ex]
\Big(1-\underset{L\in{\EuScript
L}(V_T^{v,B},H)}{\text{ess}\sup}\frac{{V_T^{v,B}}_{\mid
L}}{H}\Big) \quad &\emph{on the set} \
\left\{V_T^{v,B}\succeq_{_T}H\right\}^c .
\end{cases}
\]
\end{dfn}
\begin{rem}
Shortfall can be expressed in terms of the success function. We
have
\begin{align*}
1-\varphi_{v,B}&=0 \ \mathbf{1}_{\{V_T^{v,B}\succeq_{_T}H\}}+
\bigg(1-\underset{L\in{\EuScript L}(V_T^{v,B},H)}{\emph{ess} \
\sup}\frac{{V_T^{v,B}}_{\mid L}}{H}
\bigg) \ \mathbf{1}_{\{V_T^{v,B}\succeq_{_T}H\}^c}\\[2ex]
&=s(V_T^{v,B}).
\end{align*}
\end{rem}
Shortfall is a random variable which takes values in the interval
$[0,1]$. It is equal to $0$ if $V_T^{v,B}\succeq_{_T}H$ and it is
strictly positive if $V_T^{v,B}\nsucceq_{_T}H$. \vskip 10pt
\noindent Let $u:[0,1]\longrightarrow\mathbb{R}$ be a continuous,
non-decreasing function such that $u(0)=0$ and $u(1)<\infty$. We
regard such function as a loss function. Basing on a loss function
we define the shortfall risk of an admissible strategy as
$\mathbf{E}[u(s(V_T^{v,B}))]$. It is clear that if $v\in\Gamma(H)$
then shortfall risk is equal to $0$ for the hedging strategy,
otherwise it is positive. If for two admissible strategies $(v,B)$
and $(\bar{v},\bar{B})$ holds
$\mathbf{E}[u(s(V_T^{v,B}))]~\leq~\mathbf{E}[u(s(V_T^{\bar{v},\bar{B}}))]$
then we regard the strategy $(v,B)$ as not as risky as
$(\bar{v},\bar{B})$. If the shortfall risk of the strategy $(v,B)$
is not grater than any other, then
$(v,B)$ is called optimal or risk-minimizing.\\
Similarly to previous sections we formulate the first aspect of
risk measure problem as:
\begin{quote}
{\it For a fixed initial endowment $v_0\in\Gamma(0)$ such that
$v_0\notin\Gamma(H)$ find an admissible strategy $(v,B)$, where
$v_0\succeq_{_0}v$, such that
$\mathbf{E}[u(s(V_T^{v,B}))]\longrightarrow\min$.}\\
\end{quote}

\noindent We start with the auxiliary lemma proved in [3].
\begin{lem}
Let $X_1,X_2,...$ be a sequence of $[0,\infty)$ random variables.
There exists a sequence $\tilde{X}_n\in conv\{X_n,X_{n+1},...\}$
such that $\tilde{X}_n$ converges almost surely to a $[0,\infty]$
valued random variable $\tilde{X}$.
\end{lem}
To describe optimal strategy we start with the following theorem.
\vskip 10pt
\begin{tw}
There exists a function $\tilde{\varphi}\in\mathcal{R}$ which is a
solution of the problem
\begin{align*}
\mathbf{E}[u(1&-\varphi)]\longrightarrow\min\\[2ex]
v_0&\in\Gamma(H\varphi).
\end{align*}
\end{tw}
{\it Proof}\\
Let us denote $\mathcal{R}_0:=\{\varphi\in\mathcal{R} :
v_0\in\Gamma(H\varphi)\}$. $\mathcal{R}_0\neq\emptyset$ since
$0\in\mathcal{R}_0$. Let $\varphi_n\in\mathcal{R}_0$ be a sequence
of elements such that
$\mathbf{E}[u(1~-~\varphi_n)]~\longrightarrow~\inf_{\varphi\in\mathcal{R}_0}\mathbf{E}[u(1-\varphi)]$.
In view of lemma $6.3$ there exists a sequence
$\tilde{\varphi}_n\in conv\{\varphi_n,\varphi_{n+1,...}\}$ which
converges almost surely to $\tilde{\varphi}\in\mathcal{R}$. Since
$u(1-\tilde{\varphi}_n)\leq u(1)<\infty$, by dominated convergence
theorem we obtain
\begin{gather*}
\mathbf{E}[u(1-\tilde{\varphi})]=\lim_{n\rightarrow\infty}\mathbf{E}[u(1-\tilde{\varphi}_n)]
=\inf_{\varphi\in\mathcal{R}_0}\mathbf{E}[u(1-\varphi)].
\end{gather*}
From Fatou's lemma we have
\begin{gather*}
\mathbf{E}[\widehat{Z}_TH\tilde{\varphi}]=\mathbf{E}[\lim_{n}\widehat{Z}_TH\tilde{\varphi}_n]
\leq \underset{n}{\lim\inf} \
\mathbf{E}[\widehat{Z}_TH\tilde{\varphi}_n]\leq\widehat{Z}_0v_0\qquad\forall
Z\in\mathcal{Z}.
\end{gather*}
Hence $v_0\in\Gamma(H\tilde{\varphi})$. \hfill$\square$ \vskip10pt
\noindent The next theorem provides a description of the
risk-minimizing strategy for $v_0$. \vskip 10pt
\begin{tw}
Let $\tilde{\varphi}$ be a function from theorem $6.4$ and the
strategy $(v_0,B)$ be a hedging strategy for the modified
contingent claim $H\tilde{\varphi}$. Then $(v_0,B)$ is an optimal
strategy. Furthermore, $\tilde{\varphi}=\varphi_{v_0,B}$.
\end{tw}
{\it Proof}\\
$(v_0,B)$ is admissible since $V_T^{v_0,B}\succeq_{_T}H
\tilde{\varphi}\succeq_{_T}0$.\\
Let $(\bar{v},\bar{B})$ be any admissible strategy such that
$v_0\succeq_{_0}\bar{v}$. Then by lemma 3.6 we have
$\bar{v}\in\Gamma(H\varphi_{\bar{v},\bar{B}})$ what implies
$v_0\in\Gamma(H\varphi_{\bar{v},\bar{B}})$. From remark $6.2$ and
theorem $6.4$ we obtain
\begin{equation}
\mathbf{E}[u(s(V_T^{\bar{v},\bar{B}}))]
=\mathbf{E}[u(1-\varphi_{\bar{v},\bar{B}})]\geq\mathbf{E}[u(1-\tilde{\varphi})].
\end{equation}
Now let us consider the strategy $(v_0,B)$. Since
$V_T^{v_0,B}\succeq_{_T} H\tilde{\varphi}$, by lemma 3.7 we have:
\begin{equation}
\varphi_{v_0,B}\geq\tilde{\varphi}.
\end{equation}
Taking $(6.5.3)$ and $(6.5.4)$ into account we have
$\varphi_{v_0,B}=\tilde{\varphi}$, thus
$\mathbf{E}[u(s(V_T^{v_0,B}))]=\mathbf{E}[u(H-H\tilde{\varphi})]$
and this proves that $(v_0,B)$ is optimal. \hfill$\square$

\section{Risk measure in quantile hedging - sets with a fixed level of shortfall risk}
Assume, that we are given a number $\alpha\geq 0$. We want to
characterize strategies for which shortfall risk is not larger
than $\alpha$. This is the second aspect of risk measure problem
in quantile hedging. Our task is to provide a description of the
set $\Gamma^u_\alpha(H)$ given as
\begin{align*}
\Gamma_{\alpha}^u(H):=\bigg\{v\in\mathbb{R}^{d} : \ \text{there
exists an} &\text{admissible strategy}
\ B \\[2ex]
&\text{such that \ } \mathbf{E}\Big[u(s(V_T^{v,B}))\Big]\leq
\alpha\bigg\}.
\end{align*}
It is clear that $\Gamma^u_{\alpha_1}(H)\subseteq
\Gamma^u_{\alpha_2}(H)$ if $\alpha_1\leq\alpha_2$. Since for the
hedging strategy $(v,B)$ holds $\mathbf{E}[u(s(V_T^{v,B}))]=0$, we
conclude that set $\Gamma^u_\alpha(H)$
contains the set $\Gamma(H)=\Gamma^u_{0}(H)$ for any $\alpha\geq 0$.\\
Let us set
\begin{gather*}
\mathcal{N}:=\{\varphi\in\mathcal{R} : \quad
\mathbf{E}[u(1-\varphi)]\leq \alpha\}
\end{gather*}
The next theorem provides a description of the set
$\Gamma_\alpha^u(H)$. \vskip 10pt
\begin{tw}
The set $\Gamma_{\alpha}^u(H)$ admits the following representation
\begin{gather*}
\Gamma_{\alpha}^u(H)=\bigcup_{\varphi\in\mathcal{N}}\Gamma(H\varphi).
\end{gather*}
\end{tw}
{\it Proof}\\
This proof is similar to the proof of theorem $5.1$.\\
$\subseteq$\\
Let $v\in\Gamma_{\alpha}^u(H)$. Then there exists a strategy
$B\in\mathcal{B}$ such that $V_T^{v,B}\succeq_{_T}0$ and \\
$\mathbf{E}[u(s(V_T^{v,B}))]=
\mathbf{E}[u(1-\varphi_{v,B})]\leq\alpha$. Thus
$\varphi_{v,B}\in\mathcal{N}$ and by lemma 3.6 we have
\begin{gather*}
v\in\Gamma(H\varphi_{v,B})\subseteq\bigcup_{\varphi\in\mathcal{N}}\Gamma(H\varphi)
\end{gather*}
$\supseteq$\\
Let $v\in\bigcup_{\varphi\in\mathcal{N}}\Gamma(H\varphi)$. Then
there exists $\varphi\in\mathcal{N}$ such that
$v\in\Gamma(H\varphi)$. Let us consider the strategy $(v,B)$ which
hedges the modified contingent claim $H\varphi$. Then by lemma 3.7
we have
\begin{gather*}
V_T^{v,B}\succeq_{_T}H\varphi\quad\Longrightarrow\quad
\varphi_{v,B}\geq\varphi
\end{gather*}
and this implies
\begin{gather*}
\mathbf{E}[u(s(V_T^{v,B}))]=
\mathbf{E}[u(1-\varphi_{v,B})]\leq\mathbf{E}[u(1-\varphi)]\leq\alpha.
\end{gather*}
In effect we have $v\in\Gamma_\alpha^u(H)$. \hfill$\square$

\section{Quantile hedging under zero transaction costs}
In this section we show how the theory of F\"{o}llmer and Leukert
can be obtained. All previous sections required the {\bf EF}
condition which of course is not satisfied under zero transaction
costs. We will base on results obtained by Delbaen, Kabanov,
Valkeila [2] which are less general then results used so far, but
the condition {\bf EF} is not required there. First, we give a
short description of these results, then recall two aspects of
quantile hedging studied by F\"{o}llmer and Leukert and then show
how their theory can be obtained under zero transaction costs.\\
In cited paper we assume that transaction costs are constant in
time, given by a matrix $\Lambda$. Contingent claim is bounded
from below in the sens of partial ordering determined by the cone
$K:=M+\mathbb{R}^d_+$ thus $H\succeq c\mathbf{1}$ for some
$c\in\mathbb{R}$. $K$ is independent on $t$ and $\omega$. We
denote by $\mathcal{Q}$ the set of probability measures $Q\sim P$
such that $S_t$ follows a local martingale in respect to $Q$. We
shall need {\bf EMM} condition.
\begin{gather*}
{\bf EMM}:\qquad \mathcal{Q}\neq \emptyset.
\end{gather*}
Let $\mathcal{D}$ be the set of martingales $Z$ with $\widehat{Z}$
taking values in $K^\ast$ and bounded $\widehat{Z}_T$. Under
$\bf{EMM}$ condition we have the following description of the set
of hedging endowments :
\begin{gather*}
\Gamma(H)=\bigcap_{Z\in\mathcal{D}}\{v\in\mathbb{R}^d :
\widehat{Z}_0v\geq\mathbf{E}\widehat{Z}_TH\}.
\end{gather*}
It is left as an exercise to check that under this new description
of $\Gamma(H)$
theorems $4.1, 4.2, 5.1$, which solve our problems remain true.\\
Now take a look on a classical market model without transaction
costs. Under no-arbitrage condition the price of a scalar
contingent claim $C$ is given as
$\sup_{Q\in\mathcal{Q}}\mathbf{E}^{Q}[C]$. In the quantile hedging
problem studied by F\"{o}llmer and Leukert we consider only
admissible strategies $(x,B)$ for which the wealth process
$X_t^{x,B}\geq 0$ for all $t=0,1,...,T$. The authors use as an
effectiveness measure the success function defined as
\begin{gather*}
\varphi_{x,B}=\mathbf{1}_{\{X^{x,B}_T \geq C\}}+ \frac{X^{x,B}_T }
{C} \mathbf{1}_{\{X^{x,B}_N < C\}} .
\end{gather*}
\begin{quote}
{\underline{The first problem}}\\
{\it Let $x_0<\sup_{Q\in\mathcal{Q}}\mathbf{E}^{Q}[C]$ be a fixed
initial endowment. We search for such admissible strategy $(x,B)$,
where $x\leq x_0$, to maximize $\mathbf{E}[\varphi_{x,B}]$. We
write this as
\begin{align*}
\mathbf{E}[\varphi_{x,B}&]\longrightarrow\max\\
x&\leq x_0.
\end{align*}
}
\end{quote}

\begin{quote}
\underline{The second problem}\\
{\it Let $\varepsilon$ be a fixed number in $[0,1]$. We search for
such admissible strategy $(x,B)$ which effectiveness is not
smaller than $1-\varepsilon$ in order to minimize the initial
capital. We write this problem as
\begin{align*}
\mathbf{E}[\varphi_{x,B}&]\geq 1-\varepsilon\\
x&\longrightarrow \min.
\end{align*}
}
\end{quote}
To show that these problems can be obtained under zero transaction
costs we have to find scalar equivalents of multi-dimensional
objects on our market. Let $Y\in\mathbb{R}^d$ describes how our
wealth is allocated in stock positions on the market with
transaction costs. Now choose the $i$-th stock account to transfer
capitals from all others on it. Then the wealth of $Y$ in the
$i$-th stock is :
\begin{gather*}
Y(i):=\sum_{j=0}^{d}(1-\lambda^{ji})Y^j.
\end{gather*}
Usually $Y(i)\neq Y(j)$ for $i\neq j$, but under zero transaction
costs we have $Y(i)=Y(j)=\sum_{i=1}^dY^i$. Thus we accept the
following scalar equivalents: for the initial endowment $v$ we
take $x_v:=\sum v^i$, for the wealth process $V_t^{v,B}$ we take
$X_t^{x_v,B}:=\sum(V_t^{v,B})^i$, for the contingent
claim $H$ we take $C_H:=\sum H^i$.\\
\indent Now we show that problems of quantile hedging under zero
transaction
costs are the same as formulated by the authors for scalar equivalents.\\
First, note that
\begin{equation}
\frac{{V_T^{v,B}}_{\mid L}}{H}=\frac{\sum (V_T^{v,B})^i}{\sum H^i}
\qquad \forall L\in{\EuScript{L}}(V_T,H).
\end{equation}
For each $L\in{\EuScript{L}}(V_T,H)$ we have
$\sum\frac{{V_T^{v,B}}_{\mid L}}{H}H^i =\frac{{V_T^{v,B}}_{\mid
L}}{H}\sum H^i$ and $\sum\frac{{V_T^{v,B}}_{\mid L}}{H}H^i
=\sum({V_T^{v,B}}_{\mid L})^i$, so $\frac{{V_T^{v,B}}_{\mid
L}}{H}= \frac{\sum({V_T^{v,B}}_{\mid L})^i}{\sum H^i}$. Since the
costs are equal to zero, thus $\sum({V_T^{v,B}}_{\mid
L})^i=\sum({V_T^{v,B}})^i$ and $(8.0.5)$
holds.\\ \\
Since relation "$\succeq$" becomes a linear ordering "$\geq$" for
the sums of components, we get the equality of the success
functions.
\begin{align*}
\varphi_{v,B}&=\mathbf{1}_{\{V^{v,B}_T \succeq H\}}+
\underset{L\in{\EuScript{L}}(V_T^{v,B},H)}{\text{ess}\sup}\frac{{V_T^{v,B}}_{\mid
L}}{H}
\mathbf{1}_{\{V^{v,B}_T \succeq H\}^c}\\[2ex]
&=\mathbf{1}_{\{\sum(V^{v,B}_T)^i \geq \sum H^i \}}+
\frac{\sum(V_T^{v,B})^i}{\sum H^i}
\mathbf{1}_{\{\sum(V^{v,B}_T)^i < \sum H^i \}}\\[2ex]
&=\mathbf{1}_{\{X_T^{x_v,B}\geq C_H\}}+
\frac{X_T^{x_v,B}}{C_H}\mathbf{1}_{\{X_T^{x_v,B}< C_H\}}\\[2ex]
&=\varphi_{x_v,B}
\end{align*}
One can check that the set of the hedging endowments is of the form \\
$\Gamma(H)=\{v\in\mathbb{R}^{d}: \sum v^i\geq
\sup_{Q\in\mathcal{Q}}\mathbf{E}^{Q}[\sum H^i]\}$. Then our
problem of maximizing effectiveness
\begin{align*}
&\mathbf{E}[\varphi_{v,B}]\longrightarrow \max\\
&\quad v \preceq v_0\notin\Gamma(H)
\end{align*}
becomes
\begin{align*}
&\mathbf{E}[\varphi_{x,B}]\longrightarrow \max\\
&\quad x\leq x_{v_0}< \sup_{Q\in\mathcal{Q}}\mathbf{E}^{Q}[C_H],
\end{align*}
what is the first problem cosidered by F\"{o}llmer and Leukert.\\
Our second problem is to determine the set
$\Gamma_\varepsilon(H)$. First denote that if for $v,
\bar{v}\in\mathbb{R}^d$ holds $\sum\bar{v}^i\geq \sum v^i$ and
$v\in\Gamma_{\varepsilon}(H)$ then
$\bar{v}\in\Gamma_{\varepsilon}(H)$. For $\gamma_v:=\sum v^i$
define $\gamma:=\inf_{v\in\Gamma_{\varepsilon}(H)}\gamma_v$. If
for $v\in\mathbb{R}^d$ holds $\sum v^i\geq \gamma$ then
$v\in\Gamma_{\varepsilon}(H)$ and if $\sum v^i<\gamma$ then
$v\notin\Gamma_{\varepsilon}(H)$. Thus the set
$\Gamma_{\varepsilon}(H)$ is of the form $\Gamma_\varepsilon(H)=
\{v\in\mathbb{R}^{d} : \sum v^i\geq\gamma\}$. The problem reduces
to finding the number $\gamma$ which is the cost minimizing
capital searched by F\"{o}llmer and Leukert.
\begin{rem}
F\"{o}llmer and Leukert considered admissible strategies for which
$X_t^{x,B}\geq 0$ for each~$t=0,1,...,T$. We only require
$X_T^{x_v,B}\geq 0$, what is a generalization.
\end{rem}
{\bf Acknowledgement} I would like to thank Professor \L ukasz
Stettner for helpful discussions and Ania Stasiuk for helping with
editing this paper.
\vskip 40pt {\LARGE{\bf References}}
\begin{description}
\item{[1]}\quad
       J. Cvitanić, I. Karatzas\, \emph{On dynamic measures of risk},
       Finance and Stochastics 3 \ (1999), \ 451-482,
\item{[2]}\quad
     F. Delbaen, Yu.M. Kabanov, E. Valkeila \, \emph{Hedging under transaction costs
     in currency markets: a discrete-time model}, \ Mathematical Finance \
     12 \ (2002), \ 45-61,
\item{[3]}
     F. Delbaen, W. Schachermayer \, \emph{A general version of the fundamental
     theorem of asset pricing}, Mathematische Annalen 300 \ (1994), \
     463-520,
\item{[4]}\quad 
     H. F\"{o}llmer, P. Leukert \, \emph{Quantile Hedging},
     Finance and Stochastics 3 \ (1999), \ 251-273,
\item{[5]}\quad
     H. F\"{o}llmer, P. Leukert \, \emph{Efficient Hedging:
     Cost versus Shortfall Risk}, \ Finance and Stochastics 4 \ (2000),\
     117-146,
\item{[6]}\quad
    Yu.M. Kabanov, M. R\'{a}sonyi, Ch. Stricker\, \emph{No-arbitrage
    criteria for financial markets with efficient friction}, \
    Preprint (2001),  to appear in Finance and Stochastics,
\item{[7]}\quad
     Yu.M. Kabanov, Ch. Stricker \, \emph{The Harrison-Pliska arbitrage
     pricing theorem under \\ \qquad\qquad transaction costs}, \
     Journal of Mathematical Economics 35 \ (2001), No.2 \ 185-196,
\item{[8]}\quad
     H. Pham \, \emph{Dynamic $L^{p}$-hedging in discrete time under cone
     constrains}, SIAM J. Control Optim. 38 \ (2000), No.3 \ 665-682.
\end{description}

\end{document}